\documentstyle[11pt,aasms4]{article}

\def\ana{A. \& A.}
\def\mnras{M.N.R.A.S.}
\def\apj{Ap.J.}
\def\apjl{Ap.J. Lett.}

\begin{document}

\title{What is the True Covering Factor of Absorbing Matter in BALQSO's?}
    
\author{J.H. Krolik\altaffilmark{1}}
\author{G.M. Voit\altaffilmark{2}}

\altaffiltext{1}{Department of Physics and Astronomy, Johns Hopkins
University, Bloomberg Center, Baltimore, MD 21218; jhk@pha.jhu.edu}
\altaffiltext{2}{Space Telescope Science Institute, 3700 San Martin
Dr., Baltimore MD 21218; voit@stsci.edu}

\begin{abstract}

     Spectropolarimetry of Broad Absorption Line Quasars (BALQSO's) has
demonstrated that the geometry of the absorbing material is far from
spherically symmetric.  Calculations of accretion disk spectra
suggest that the intrinsic radiation pattern of quasars is also anisotropic.
Because the quasar counts distribution is very steep, if the orientations of
these two anisotropies are correlated, optical flux-limited samples would
be very strongly biassed with respect to discovery of BALQSO's.  In particular,
currently favored models suggest that BALQSO's may be much {\it more}
common than in the samples.  If so, the intrinsic covering fraction of
absorbing matter must be relatively large.  Numerous consequences follow,
including prediction of a new population of hard X-ray sources (as may have
already been seen by ASCA), and a possible explanation of the anomalously
strong NV~1240 lines seen in many high redshift quasars.

\end{abstract}


\section{Asymmetric Absorption and Anisotropic Intrinsic Radiation}

     Only about 10\% of quasars found in optical flux-limited samples
exhibit broad absorption lines, resonance line absorption troughs extending
$\sim 0.1c$ to the blue of the emission line centers (Weymann et al. 1991).
Because in most other respects Broad Absorption Line Quasars (BALQSO's)
are quite similar to ordinary quasars, it is generally thought that
rapidly-moving absorbing matter exists in {\it all} quasars, but occupies
only a fraction of the solid angle around them (Weymann et al. 1991).
The fact that the remaining light in the troughs is often strongly polarized
indicates that the obscuration is not randomly distributed around the quasar
in a statistically spherically symmetric manner, but is instead strongly
aspherical.  Equatorial wedges are the most popular geometry used to
explain the observations (Goodrich \& Miller 1995; Cohen et al. 1995;
Brotherton et al. 1997).

     Quasars are also likely to be anisotropic in their intrinsic continuum
emission.  The most plausible source for the photons they produce in the
rest-frame optical and ultraviolet bands is an accretion disk, although
detailed models for this emission are not altogether successful (see Blaes
1998 and Krolik 1998 for reviews).
While the specific character of the intrinsic angular radiation pattern
is highly model-dependent, in virtually every model the anisotropy is
significant.  Typically, the wavelength range around 1000 -- 2000\AA\  is
limb-darkened, while shorter wavelengths, produced closer to the
black hole, are limb-brightened (Sun \& Malkan 1989, Laor \& Netzer 1989,
Agol 1997).

     If the orientations of the intrinsic radiation pattern and the
absorbing matter are correlated, optical flux-limited samples will be
biassed with respect to the detection of BALQSO's (cf. Goodrich 1997,
who suggested that there is bias due to extra attenuation on the lines
of sight through the absorbing matter).  This effect is strengthened by the
steepness of the quasar counts distribution.

   For example, suppose that the differential source count distribution
{\it if quasars radiated isotropically} were $dN/dF_* = K F_*^{-\alpha}$,
where $N(>F_*)$ is the number of quasars per sky solid angle with flux
greater than isotropic flux $F_*$.
Then the distribution per cosine of the viewing angle is
$\partial^2 N/\partial F_*\partial\mu = (K/2)F_*^{-\alpha}$.  If we now
allow for anisotropy by setting $F_{obs} = F_* f(\mu)$ (implicitly
assuming that the angular radiation pattern is independent of all other
quasar characteristics such as luminosity or redshift), the differential
source count distribution that we would actually observe is
\begin{equation}
{\partial N \over \partial F_{obs}} = 
\int_{-1}^{1} \, d\mu \, {\partial^2 N \over \partial F_{obs}\partial\mu} = \int_{-1}^{1} \, d\mu \, {\partial^2 N \over \partial F_{*}\partial\mu}
{\partial F_* \over \partial F_{obs}} = 
{K \over 2} F_{obs}^{-\alpha}\int_{-1}^{1} \, d\mu \,  f^{\alpha-1} (\mu).
\end{equation}

  For optical fluxes in the neighborhood of $m_B = 16$ -- 17,
$\alpha \simeq 3.25$ (Hartwick \& Schade 1990).
Since a quasar is only called a BALQSO if it is found within an equatorial
wedge defined by $|\mu| \leq \mu_*$, the BAL fraction in a flux-limited
sample is
\begin{equation}
f_{BAL} = 
{\int_{0}^{\mu_*} \, d\mu \, \partial^2 N/\partial F_{obs}\partial\mu \over
\int_{0}^{1} \, d\mu \, \partial^2 N/\partial F_{obs}\partial\mu} =
{\int_{0}^{\mu_*} \, d\mu \, f^{\alpha -1} (\mu) \over
\int_{0}^{1} \, d\mu \, f^{\alpha -1} (\mu)} .
\end{equation}
Thus, if quasars are dimmer in the BAL direction, their population fraction
at fixed {\it observed} flux is diminished because they are submerged in the larger population of intrinsically fainter non-BAL quasars.

    As already remarked, we cannot compute $f(\mu)$ with great confidence.
However, two plausible guesses will illustrate the possible magnitude of
its effect.  The first guess is the simplest possible model: that the local
limb-darkening is given by the form appropriate to a gray atmosphere in
the Eddington approximation (Mihalas 1978), but that the
integrated flux is further multiplied by a factor of $\mu$ because the
radiation comes from a flat disk.  Then $f_1(\mu) = |\mu|(1 + 1.5|\mu|)$.
As a foil to this form, we use a rough analytic fit to detailed
radiation transfer calculations which also include a complete treatment of
general relativistic corrections (E. Agol, private communication).  This
analytic fit is $f_2 (|\mu|) \simeq 0.23 + 1.54|\mu|$.

    The predicted BAL fractions are
shown in Fig. 1.  Generically, {\it any} significant limb-darkening in
the direction of the absorbing matter leads to a very strong bias against
the discovery of BALQSO's.  In order to find $\sim 10\%$ BALQSO's in an
optical flux-limited sample of limb-darkened sources, the true covering
fraction $\mu_* \simeq 0.5$.
When $\mu_*$ exceeds $\simeq 0.5$, the fraction of BALQSO's discovered in
a flux-limited sample rapidly approaches unity as the limb-darkening weakens
and the fraction of non-BALQSO's falls.  Conversely, if the absorbing
matter is correlated with the {\it bright} direction, the true fraction
might be substantially less than the fraction inferred from flux-limited
samples.

\vbox to 4.5in{
\plotfiddle{"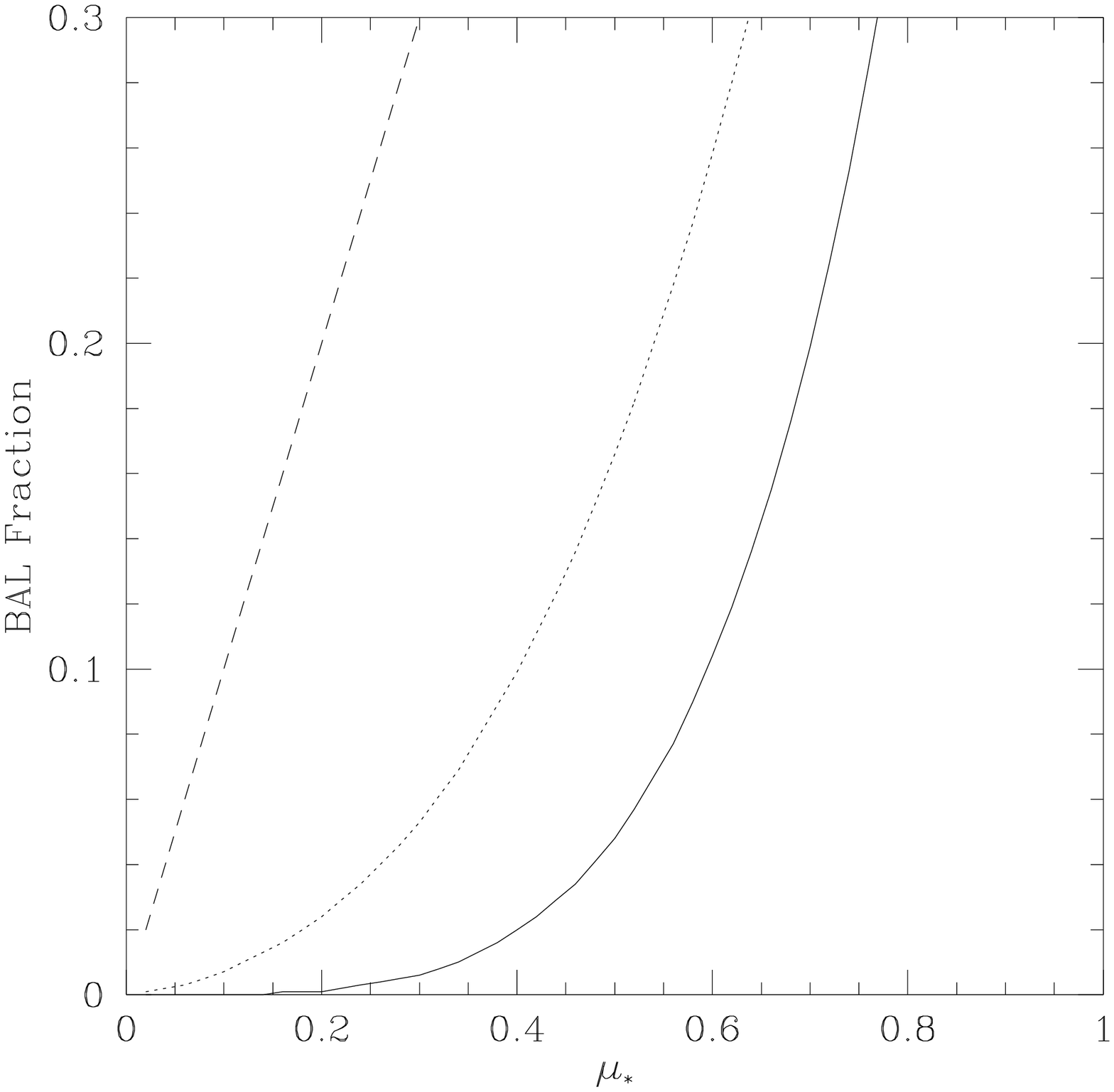"}{2.5in}{0}{50}{50}{-20}{-100}
\parbox{4.0in}{
\vbox to 180pt{\vfill}
\small\baselineskip 9pt
{\sc Fig.}~1.---The fraction of BALQSO's that would apear in an optical flux-limited
sample for three different limb-darkening laws.  The solid curve is the
prediction of gray atmosphere limb-darkening on a disk, the dotted
curve the approximation to detailed disk models, and the dashed line gives
$f_{BAL}$ if the continuum were radiated isotropically.
}}

    This picture also has one more corollary regarding the BAL fraction:
if the absorbing matter is
in the dim direction, the very faintest quasars (i.e. those at flux levels
where $\alpha$ falls below unity) should be disproportionately BALQSO's.

\section{Consequences of Large Covering Fractions}

     If the true covering fraction of absorbing matter is as large as would
be predicted if the dim continuum direction is correlated with the
direction in which absorption is found, our picture of quasars changes
substantially.  The outward flow of matter moving at $\sim 0.1c$ is no longer
a small effect present in only a limited solid angle; it is, rather, a major
element of the system, carrying substantially greater amounts of momentum
and energy.  Moreover, the velocity is no longer directed only
within a small range of directions close to the surface of the accretion disk;
there are fluid elements moving at very substantial angles relative to the
disk surface.  These considerations are likely to change our approach in
dynamical studies of the BAL material.

    Beyond these dynamical issues, there are also a number of phenomenological
consequences to this changed point of view.

\subsection{A new AGN X-ray population}

    For only one BALQSO, PHL 5200, has an X-ray spectrum of good enough quality
to determine an absorbing column been obtained; in this instance it is
$\simeq 1 \times 10^{23}$~cm$^{-2}$ (Mathur et al. 1995).  However, it has
been possible to show statistically that most BALQSO's are relatively weak
in the soft X-ray band relative to their optical/ultraviolet flux.  Interpreted
as the result of absorption, the depression corresponds to columns $\sim
10^{22}$~cm$^{-2}$ (Green \& Mathur 1996).

    If the fraction of BALQSO's is nearer 0.5 than 0.1, and absorbing columns
$\sim 10^{22}$ -- $10^{23}$~cm$^{-2}$ thick are a generic feature of BALQSO's,
the number of quasars turned up in X-ray surveys sensitive to rest-frame photon
energies greater than a few keV should increase by roughly a factor of two
relative to the number found by soft X-ray surveys.  Anisotropy in the X-ray
emission is likely to be rather weaker than in the ultraviolet, but if present
could modify this prediction.  These new quasars should have
hard X-ray spectra at rest-frame energies 1 -- 5~keV
(as a result of the absorption), and most of them should also show BAL features
in their rest-frame ultraviolet spectra.

   Interestingly, the ASCA source
count distribution shows exactly this effect: The
number of sources seen by ASCA with fluxes
$\sim 10^{-13}$ -- $10^{-12}$~erg~cm$^{-2}$~s$^{-1}$ in the 2 -- 10~keV band
is several times as great as the number that would be predicted on the
basis of the corresponding Rosat counts if the sources had conventional
quasar spectra (Cagnoni et al. 1997).  Some of these may be AGN obscured
by dusty tori as in Seyfert galaxy unification models (Madau et al. 1994;
Comastri et al. 1995), but others may be BALQSO's.

\subsection{The radio-loudness of BALQSO's}

  As already discussed by Goodrich (1997), a reduction in the optical flux
of BALQSO's {\it increases} their ratio of radio to optical flux.  He found
that the (rough) statistics of known BALQSO's indicated that their optical
flux had been depressed relative to the mean of all quasars by a factor of
3 -- 5.

   More precisely, we wish to compute the ratio of radio flux (preferably
extended radio flux, so that it is isotropic) to observed optical
flux for quasars found in optically-selected samples, i.e. at fixed
observed optical flux.  If the radio flux $F_r = R F_*$, then for a single
quasar, the ratio of observed radio flux to observed optical flux is
$F_r/F_{obs} = R/f(\mu)$.
Averaging over viewing angle at fixed $F_{obs}$, the mean radio to optical
ratio for BALQSOs is 
\begin{equation}
\langle F_r / F_{obs}\rangle_{BAL} = 
{\int_{0}^{\mu_*} \, d\mu \, [R/f(\mu)]\partial^2 N /
\partial F_{obs} \partial \mu 
\over \int_{0}^{\mu_*} \, d\mu \, \partial^2 N /
\partial F_{obs} \partial \mu} = 
R {\int_{0}^{\mu_*} \, d\mu \, f^{\alpha-2} (\mu) \over 
\int_{0}^{\mu_*} \, d\mu \, f^{\alpha-1} (\mu)},
\end{equation}
so that 
\begin{equation}
{\langle F_r/F_{obs} \rangle_{BAL} \over \langle F_r/F_{obs} \rangle_{all}} = 
{1 \over f_{BAL}}{\int_{0}^{\mu_*} \, d\mu f^{\alpha-2} (\mu) 
\over \int_{0}^{1} \, d\mu f^{\alpha-2} (\mu)}.
\end{equation}
If either of the two models for $f(\mu)$ used in \S 1 is assumed with a
value of $\mu_*$ that gives an observed BAL fraction of $\sim 0.1$,
$\langle F_r/F_{obs}\rangle_{BAL}$ is enhanced relative to the mean ratio
for all quasars by $\simeq 2$.

\subsection{Scattering contributions to emission lines}

     Many people have thought about possible contributions to the broad
emission lines due to resonance scattering in BAL gas
(Surdej \& Hutsemekers 1987; Turnshek 1988; Hamann et al. 1993; Hamann \&
Korista 1996).  Indeed, some (e.g. Turnshek 1988 and Hamann et al. 1993) have
tried to limit the covering
fraction by constraints on these contributions.  However,
several special features arise in this model, which change the
character of these arguments: the large covering fraction
enhances the magnitude of the scattering contributions, although the
limb-darkening partially counteracts this; and the width of the
scattered line is diminished both by the preference for discovering
face-on quasars, and by the relatively greater equivalent width of
the intrinsic emission line as seen by the scattering gas.

\subsubsection{Emission line equivalent width}

      If we consider an isolated resonance line (e.g. CIV~1549), the
limb-darkening by itself can modulate the equivalent width of the
intrinsic emission line by factors of several (as would similarly occur
for isolated non-resonance lines such as H$\beta$ or CIII]~1909).
However, the combination of scattering and absorption diminishes the
range of observed emission
equivalent widths.  Scattering increases the equivalent width of
emission lines seen from the polar direction, which would otherwise be
the direction in which the equivalent width was least, while absorption,
of course, removes line flux from the equatorial direction.

    The quantitative character of these effects depends strongly on details of
the geometry and kinematics.  It matters whether the emission region
is very small compared to the absorption region, or nearly fills it.  The
scattered line profile also depends on the velocity gradients within the
scattering gas, and on the direction of the flow (helical outflow, as
predicted by Murray et al. 1995, has different properties from radial, for
example).

  To qualitatively explore scattering effects, consider an extremely
simplified model: Suppose that the emission line region is far inside the
absorption region, the velocity of the absorbing gas is purely radial, and
the absorbing gas is black across some range of velocities $\beta_{min}
\leq \beta \leq \beta_{max}$, but transparent otherwise.  With these
simplifying assumptions, the scattered luminosity in the direction $\hat n$ is
\begin{equation}
{dL_\nu \over d\Omega} = \int_{\beta_{min}}^{\beta_{max}} \, d\beta \,
\int_{-\mu_*}^{+\mu_*} \, d\mu^{\prime} \, \int_{0}^{2\pi} \, 
d\phi^{\prime} \, 
g(\hat n,\hat n^{\prime}) {d L_\nu^{\prime} \over d\Omega^{\prime}}
\delta (\beta - \beta_*),
\end{equation}
where primed quantities refer to the intrinsic spectrum.  Photons of
initial frequency $\nu^{\prime}$ and direction $\hat n^{\prime}$ are
scattered with probability $g(\hat n, \hat n^{\prime})$ into direction
$\hat n$ and frequency $\nu$ by material with
speed $\beta_*$, defined by the pair of equations
\begin{eqnarray}
\nu^{\prime} &= \nu_o \sqrt{{1 + \beta_* \over 1 - \beta_*}} \\
\nu^{\prime} &= \nu{1 - \beta_* \hat n^{\prime} \cdot \hat n \over 1 - \beta_*},
\end{eqnarray}
for rest-frame resonance frequency $\nu_o$.

When $\mu_* \simeq 0.5$, $\beta_{min} \simeq 0$, and
$\beta_{max} \simeq 0.05$ -- 0.1, the observed emission equivalent
width of an isolated resonance line seen in a non-BAL direction
is typically enhanced by a few tens of
percent relative to what it would be without scattering, and ranges over
roughly a factor of two from the polar to the equatorial directions.  Part
of this enhancement in equivalent width is due to low broad wings.  Depending
on how the equivalent width is measured, in practise only a portion of
these wings may actually be counted towards the emission line equivalent
width, with the rest absorbed into the apparent continuum.
Interestingly, if the maximum velocity in the absorbing gas is
$\sim 0.1c$, the extent of the wings is comparable to the extent of the
``very broad line" component identified by Brotherton et al. (1994).

  If attention is restricted to the red half of the emission line, this model
predicts a greater contrast between BAL and non-BAL quasars.  The emission
equivalent width in BALQSOs should be some 50\%
stronger, on average, than similar lines in unabsorbed quasars.
Weymann et al. (1991) found no strong differences between the emission-line 
properties of quasars with and without broad absorption; however, the 
dispersion of properties in each sample is large.  An additional effect
that might mask possible differences is that the intrinsic emission might 
be blueshifted relative to the systemic velocity of the quasar, so that
non-BALQSO's appear to have greater red-side flux.  Investigation of such
effects will have to await more sophisticated modelling of line profiles 
in limb darkened BALQSOs.

\subsubsection{Ratio of absorption to emission equivalent width}

    The ratio of the absorption equivalent width to the emission equivalent
width is sometimes taken as an indicator of the intrinsic covering fraction
because scattering by itself would result in these being equal if the
covering fraction were unity (e.g. Hamann et al. 1993).  It is, of course,
reduced by any true emission.  The observed range of this ratio in BALQSO's is $\simeq 5$ -- 0.5, with a median value
$\simeq 2$ (Hamann et al. 1993).  In the simple model defined in
the preceding paragraph, this ratio also depends on both the intrinsic
emission equivalent width and the range of velocities in the absorbing
matter, decreasing for greater intrinsic equivalent width and increasing
for increasing $\beta_{max}$.  For $\mu_* \simeq 0.5$, it is 2 -- 4 when
$\beta_{max} \simeq 0.1$ and $W_\lambda/\lambda \simeq 0.01$ -- 0.015.

\subsubsection{Emission line profiles}

   Hamann et al. (1993) argued that if the covering fraction of scattering
gas were large, the resulting emission line profile (in the non-BAL view)
would be too broad.  However, if the intrinsic emission line flux is originally 
isotropic while the continuum radiation is limb-darkened, this story
changes.  A wind in an equatorial wedge with a covering factor of
0.5 can intercept half of the isotropic line photons while intercepting
a much smaller fraction of the limb-darkened continuum.  The scattered
C~IV profile that results is then not much broader than the original 
line because the contribution from scattered line photons dominates
that from scattered continuum photons. Scattering in the radial outflow 
can only {\em reduce} a photon's frequency in the rest frame of the 
photon source.  

We have explored this effect using the very simple model of scattering
described above.  Suppose, for example, that
the absorption profile is generated by
pure isotropic scattering, $\beta_{max} = 0.05$, and $\mu_* = 0.5$.
Let us further assume that the continuum limb-darkening is given by
$f_2 (\mu)$, and that the intrinic C~IV emission line,
emitted by a point source at the center, is Gaussian with a dispersion 
of $0.01 c$ and peak-to-continuum ratio of 1.5 when viewed along the 
pole.  The resulting C~IV profile, when viewed over a variety of
non-BALQSO lines of sight, has a FWHM of $0.015 c$, a peak-to-continuum
ratio $\sim 2$, and a shape parameter $S \equiv (FW_{\rm 1/4 \, max} - 
FW_{\rm 3/4 \, max})/FWHM \sim 1.0$.  All these numbers compare favorably
with the C~IV line-shape parameters measured by Wills et al. (1993)
for quasars in the same redshift range as the bulk of the Weymann et al.
(1991) sample.  The red sides of the model C~IV BALQSO profiles have
similar shapes but somewhat higher peak-to-continuum ratios ($\sim 2.5$)
because the continuum is substantially limb darkened.

\subsubsection{Enhancement of NV~1240}

   The pair of lines L$\alpha$ and NV~1240 is a special case.  Because the
absorbing gas sees the intrinsic emission {\it red-shifted}, and
the intrinsically strong L$\alpha$ emission line
is blueward of NV~1240 by only $\simeq 6000$~km~s$^{-1}$, the flux available
to scatter in the NV feature is substantially magnified relative to the
pure continuum.  For this reason, scattering can increase the NV~1240
emission equivalent width by factors of several (Surdej \& Hutsemakers 1987),
provided that the covering factor $ > 0.3$ (Hamann et al. 1993).

    To quantify this effect, we again use the simple model already described. 
In this case, because NV~1240 and Ly$\alpha$ are always blended, one cannot
measure the equivalent width of the observed NV~1240 line by integrating
across the profile after subtracting off a fixed continuum level.  Instead,
we defined the flux in each of the two lines by a least squares fit to
a model of two Gaussians, one centered at 1240\AA, the other at 1216\AA.
Specific numbers for the relative strength of the apparent NV emission
due only to scattering depend on parameters, but only weakly.  For
$\mu_* = 0.5$, limb-darkening law $f_2(\mu)$, $\beta_{min} = 0$,
$\beta_{max} \simeq 0.05$ -- 0.1, and L$\alpha$ equivalent width from
40 -- 140\AA (a range taken from Baldwin et al. 1989), the scattered
NV~1240 flux is 0.4 -- $0.5 \times$ the L$\alpha$ flux.  This ratio---with
{\it no} contribution from intrinsic NV emission---is right in the middle
of the range found by Baldwin et al. (1989) for their sample of quasars
with $z \simeq 1.7$ -- 2.5.

    Strong enhancement of the NV~1240 line is of particular interest in
view of the fact that its flux, relative to both CIV~1549 and HeII~1640,
increases on average by a factor of 2 -- 3 comparing AGN at $z \sim 0.1$ with those at $z \simeq 2$ -- 4 (Hamann \& Ferland 1992, 1993).  If the incidence
of BALQSO's also increases sharply with increasing redshift (as anecdotal
evidence suggests), the scattering contribution may explain this systematic
change in line ratios.  Further support to this idea that the strength
of NV~1240 is due to scattering in the BAL gas is given by the principal
component analysis of quasar spectra by Francis et al. (1992),
which showed that broad absorption is related to excess N~V emission and
broader C~IV emission.

\section{Conclusions}

    We have shown that if anisotropy in the continuum emission of quasars
is correlated with the direction towards broad absorption line gas,
optical flux-limited samples of quasars are severely biassed with respect
to the discovery of broad absorption line quasars.  If, as the simplest
models suggest, quasar continua are limb-darkened at the selection
wavelengths, BALQSO's are severely underrepresented in such samples.  The
inferred covering fraction of broad absorption gas could well be nearer
$\simeq 0.5$ than the usual estimate of $\simeq 0.1$.

     Such large covering fractions of absorbing gas would have numerous
consequences: Hard X-ray surveys may discover roughly twice as many
quasars as would be predicted on the basis of soft X-ray surveys;
the ASCA survey may already have done just this.  The fact that BALQSO's are
somewhat stronger radio sources than the average radio-quiet quasar
would also be explained.  In addition, resonance scattering in the
broad absorption gas could contribute to the emission lines.  NV~1240,
because it is just 6000~km~s$^{-1}$ redward of Ly$\alpha$, would be
strongly enhanced, providing an alternative explanation
[i.e. other than elevated N abundance: Hamann \& Ferland (1992, 1993)]
for the strength of this line in high redshift quasars.

\acknowledgments

This work was partially supported by NASA Grant NAG5-3929 and NSF Grant
AST-9616922.

\end{document}